\documentclass[12pt]{iopart}

\usepackage{graphicx}

\begin{document}

\title[Graphenylene nanoribbons: electronic, optical and thermoelectric properties]{Graphenylene nanoribbons: electronic, optical and thermoelectric properties from first-principles calculations}

\author{R.M. Meftakhutdinov$^1$, R.T. Sibatov$^{1,2}$, A.I. Kochaev$^1$}

\address{$^1$Ulyanovsk State University, Ulyanovsk, Russia\\
$^2$Institute of Nanotechnology of Microelectronics of the Russian Academy of Sciences (INME RAS), Moscow, Russia}
\ead{ren\_sib@bk.ru}
\vspace{10pt}

\begin{abstract}
	
Recently synthesized two-dimensional graphene-like material referred to as graphenylene is a semiconductor with a narrow direct bandgap
 that holds great promise for nanoelectronic applications. 
The significant bandgap increase can be provided by the strain applied to graphenylene crystal lattice or by using nanoribbons instead of extended layers. 
In this paper, we present the systematic study of the electronic, optical and thermoelectric properties of graphenylene nanoribbons using calculations based on the density functional theory. 
Estimating the binding energies, we substantiate the stability of nanoribbons with zigzag and armchair edges passivated by hydrogen atoms.
Electronic spectra indicate that all considered structures could be classified as direct bandgap semiconductors. The absorption coefficient, optical conductivity, and complex refractive index are calculated by means of the first-principles methods and the Kubo-Greenwood formula. It has been shown that graphenylene ribbons effectively absorb visible-range electromagnetic waves. Due to this absorption the conductivity is noticeably increased in this range. The transport coefficients and thermoelectric figure of merit are calculated  by the nonequilibrium Green functions method. Summarizing the results, we discuss the possible use of graphenylene films and nanoribbons in nanoelectronic devices. 

\end{abstract}

%
%
%
%
%

\section{Introduction}

Advances in nanoelectronics, hydrogen energy storage, and biotechnology are closely related to the development of novel nanoscale materials and structures exhibiting unique physical properties. 
Recently, the attention of material scientists has been actively attracted to the $sp^2$ carbon nanoallotrope with unique geometry. It is referred as to 4-6 carbophene, biphenylene carbon or graphenylene. For the first time, a hypothetical  two-dimensional material with graphenylene structure has been introduced in Ref.~\cite{Balaban1968}.

\begin{figure}[!h]
	\centering
	\includegraphics[width=0.65\linewidth]{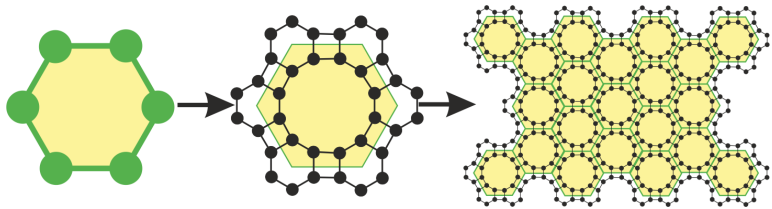}
	\caption{Geometry of the graphenylene crystal lattice composed of hexagonal graphene cells.}
	\label{ko4_fig1_} 
\end{figure}

Graphenylene is a graphene-like material in the sense that its infinite representation can be considered as a two-dimensional crystal formed by hexagonal cells with nodes containing a hexagon of carbon atoms. The geometric structure of graphenylene is illustrated in Fig.~\ref{ko4_fig1_}. The presence  of periodically arranged pores with a diameter of 3.2~\AA\ in the graphenylene plane opens up a new route to the appearance of effective methods for hydrogen separation~\cite{Song2013a}.

For a long time, graphenylene has been considered as hypothetical two-dimensional material.  However, in 2017, Qi-Shi Du et al.~\cite{Du2017} succeeded in synthesizing graphenylene layers by the dehydration and polymerization of 1,3,5-trihydroxybenzene referring the obtained material to as 4-6 carbophene. Due to distorted $sp^2$ orbitals, this two-dimensional material is less stable than graphene, but, nevertheless, it can exist in the environment under normal conditions.

The optimization and calculation of the electronic structure of graphenylene have been reported in Refs.~\cite{Brunetto2012, Enyashin2011, Brazhe2016, Song2013a}. In these studies, the obtained bandgap values $E_g$ sufficiently differ from each other. Since the bandgap is rather narrow, $E_g$ value is extremely sensitive to atom positions in the optimized structure and strongly dependent on the calculation method. By means of the DFT calculations, Song et al.~\cite{Song2013a} obtained $E_g = 0.025$~eV. In our work, the value $ E_g = 0.051$~eV is derived using the DFT method implemented in the QuantumATK software \cite{Smidstrup2017} (the method is described below).

\begin{figure}[tbh]
	\centering
	\includegraphics[width=0.7\linewidth]{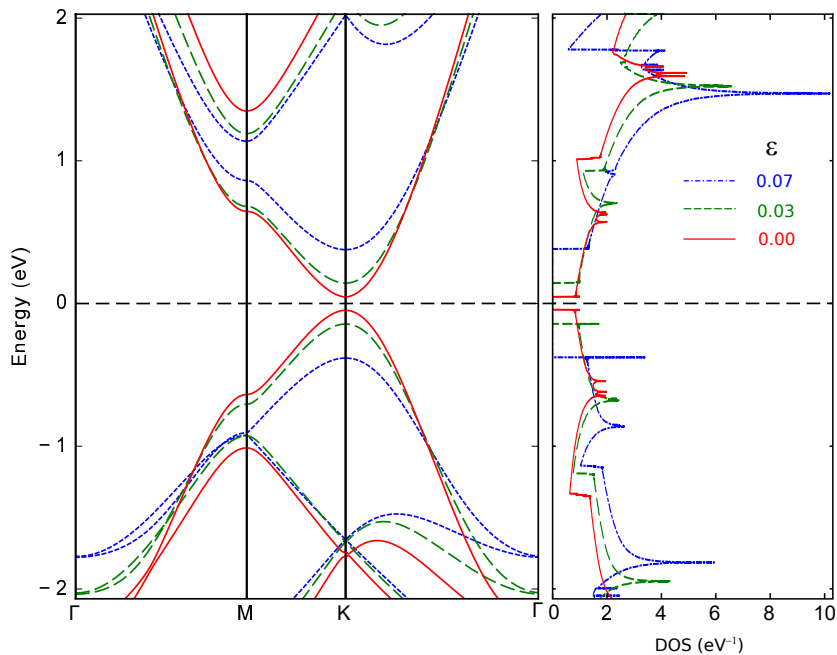}\\
	\caption{The band structure of free and stressed graphenylene sheets. The dashed line shows the Fermi level.  DFT with GGA-PBEsol functional the 9$\times$9$\times$1 grid Monkhorst-Pack grid (see details in the text below).}
	\label{fig:bg_grphenylen}
\end{figure}

The calculated bandstructure (Fig.~\ref{fig:bg_grphenylen}) demonstrates that graphenylene is a two-valley direct-gap semiconductor with a narrow bandgap equal to $ E_{g} = 0.051$~eV at $K$ point. The second valley is located at $M$ point, and the difference between the conduction band minimum and the valence band maximum at $M$ is~1.283~eV.

The bandgap of graphenylene is easily controlled by tensile strain. Figure~\ref{fig:bg_grphenylen} shows the bandstructure of graphenylene under tensile deformations $\varepsilon=0.03$ and 0.07. 

Koch et al.~\cite{Koch2015} investigated the electronic properties of graphenylene nanotubes. By DFT calculations, they showed that graphenylene nanotubes with ``zigzag'' edges are semiconductors with a small bandgap values, while narrow ``armchair'' tubes exhibit metallic properties and become semiconducting for increased diameters.

One of the current research directions in the graphenylene topic is related to the replacement of carbon atoms in the lattice with atoms of other chemical elements completely~\cite{Zhang2017}, or partially~\cite{Freitas2018}. These changes in the structure do not lead to a significant deterioration in thermodynamic stability, while the electronic properties become more diverse~\cite{Zhang2017, Freitas2018}.

The aim of this work is the systematical study of electronic, optical, and thermoelectric properties of graphenylene nanoribbons (GRNR) by means of the first-principles methods. 

\section{Materials}

\begin{figure}[h!]
	\centering
	\includegraphics[width=0.95\textwidth]{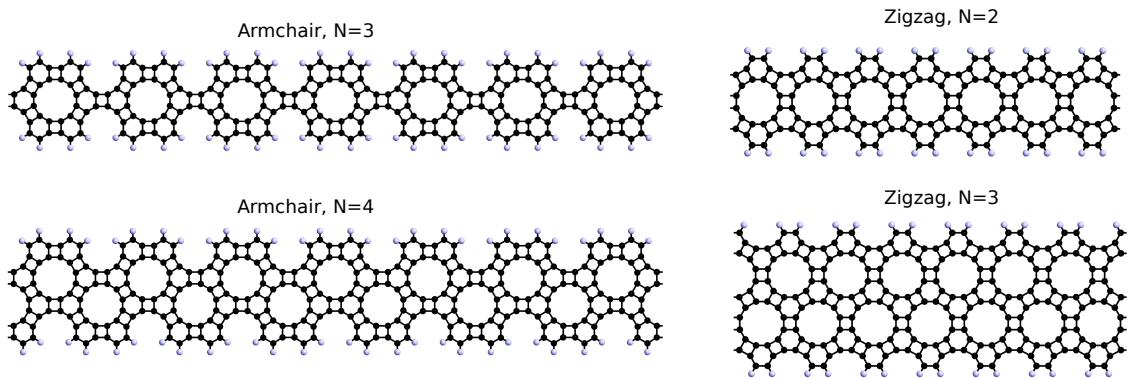}
	\caption{Crystal structures of some graphenylene nanoribbons with edges of ``armchair'' (left) and ``zigzag'' (right) type.}
	\label{ko4_fig2}
\end{figure}

In this work, graphenylene nanoribbons (GRNR) with zigzag and armchair edges passivated by hydrogen atoms  are considered.
Similar to graphene ribbons, the graphenylene ones can be distinguished by the number $N$ of zigzag chains for ZGRNR and the number $N$ of dimeric lines  for AGRNR. In this context, the abbreviations $N$-ZGRNR and $N$-AGRNR can be introduced. Figure~\ref{ko4_fig2} shows the crystal structures of some graphenylene nanoribbons with ``zigzag'' and ``armchair'' edges. To exclude dangling bonds in the system, the nanoribbons are passivated by hydrogen atoms.

\section{Thermodynamic stability}

The results presented in this paper are obtained by means of  DFT and NEGF methods implemented in the Quantum ATK~2018.06-SP1-1 package~\cite{Smidstrup2017}. In all cases, the effect of ionic nuclei is described by the pseudopotentials PseudoDojo~\cite{Setten2018}. Electronic wave functions are presented in the basis of plane waves with a cutoff energy of 500 eV. Electron exchange and correlation effects are described in the framework of the Purdue-Burke-Ernserhof approximation generalized to the case of solids (GGA-PBEsol)~\cite{Csonka2009}. To generate $k$ points in the Brillouin zone, the Monkhorst-Pack method~\cite{Monkhorst1976} with the  1$\times$1$\times$24 grid is used for GRNR. The structure is optimized while atomic forces exceeds 0.01 eV/\AA.  To prevent the influence of boundary conditions in the $z$-direction perpendicular to the sheet, we take the value of (80~\AA) of the cell parameter along $z$, so that it significantly exceeds the possible bond length (10 \AA).

The binding energy per atom is calculated as follows:
\begin{displaymath}
E_b=\frac{E_t-NE}{N},
\end{displaymath}
where $E$ is the total energy of an isolated atom, $N$ is the number of atoms in the translated unit cell, and $E_t$ is the total energy of this cell. The fragmentation energy of the system is considered as the zero energy level. 
For graphene, this formula gives the binding energy of $-7.83$~eV/atom, demonstrating good agreement with the results obtained by other authors~\cite{Koskinen2008}.
The equilibrium values of bond lengths and bond energies per atom are summarized in table~\ref{tab:stab}.

\begin{table}[h!]
	\caption{Optimized parameters and bandgaps of armchair and zigzag graphenylene nanoribbons}
	\label{tab:stab} 
	{\footnotesize
	\begin{center}
		\begin{tabular}{llllllc}
			\hline
			Type of nanoribbon & $E_b$, eV/atom & $l_a$, \AA & $l_b$, \AA & $l_c$, \AA & $E_{g}$, eV \\
			\hline
			A3  & --6.86 & 1.492 & 1.449 & 1.373 & 0.962 & \\
			A4  & --7.02 & 1.488 & 1.464 & 1.374 & 0.710 & \\
			A5  & --7.12 & 1.486 & 1.457 & 1.368 & 0.458 & \\
			A6  & --7.19 & 1.483 & 1.468 & 1.372 & 0.346 & \\
			A7  & --7.24 & 1.469 & 1.481 & 1.366 & 0.320 & \\
			A11 & --7.36 & 1.472 & 1.478 & 1.367 & 0.144 & \\
			\hline
			Z2  & --7.02 & 1.480 & 1.448 & 1.384 & 0.786 & \\
			Z3  & --7.19 & 1.479 & 1.464 & 1.375 & 0.430 & \\
			Z4  & --7.28 & 1.470 & 1.460 & 1.375 & 0.294 & \\
			Z5  & --7.34 & 1.473 & 1.468 & 1.395 & 0.177 & \\
			Z6  & --7.37 & 1.472 & 1.463 & 1.374 & 0.149 & \\
			Z11 & --7.47 & 1.477 & 1.474 & 1.371 & 0.078 & \\
			\hline
		\end{tabular}
	\end{center} 
}
\end{table}

Table~\ref{tab:stab} shows that the bond lengths of optimized graphenylene nanoribbons range from 1.366 to 1.492 \AA. As the width increases, the binding energy $E_b$ grows and tends to the value corresponding to the graphenylene sheet (7.58 eV/atom).  The presented $E_b$ values indicate a slightly worse thermodynamic stability of the graphenylene nanoribbons in comparison with the graphene ribbons, but comparable with values typical for 2D hexagonal boron nitride structures~\cite{Li2017}.

\section{Electronic Properties}

The electronic structure of nanoribbons is calculated by the DFT method with the generalized gradient approximation (GGA) and the PBE-sol exchange-correlation functional. The $k$-point grid 1$\times$1$\times$24 in the first Brillouin zone is generated by the Monhorst-Pack method.

\begin{figure}[p]
	\centering
	\includegraphics[width=0.8\linewidth]{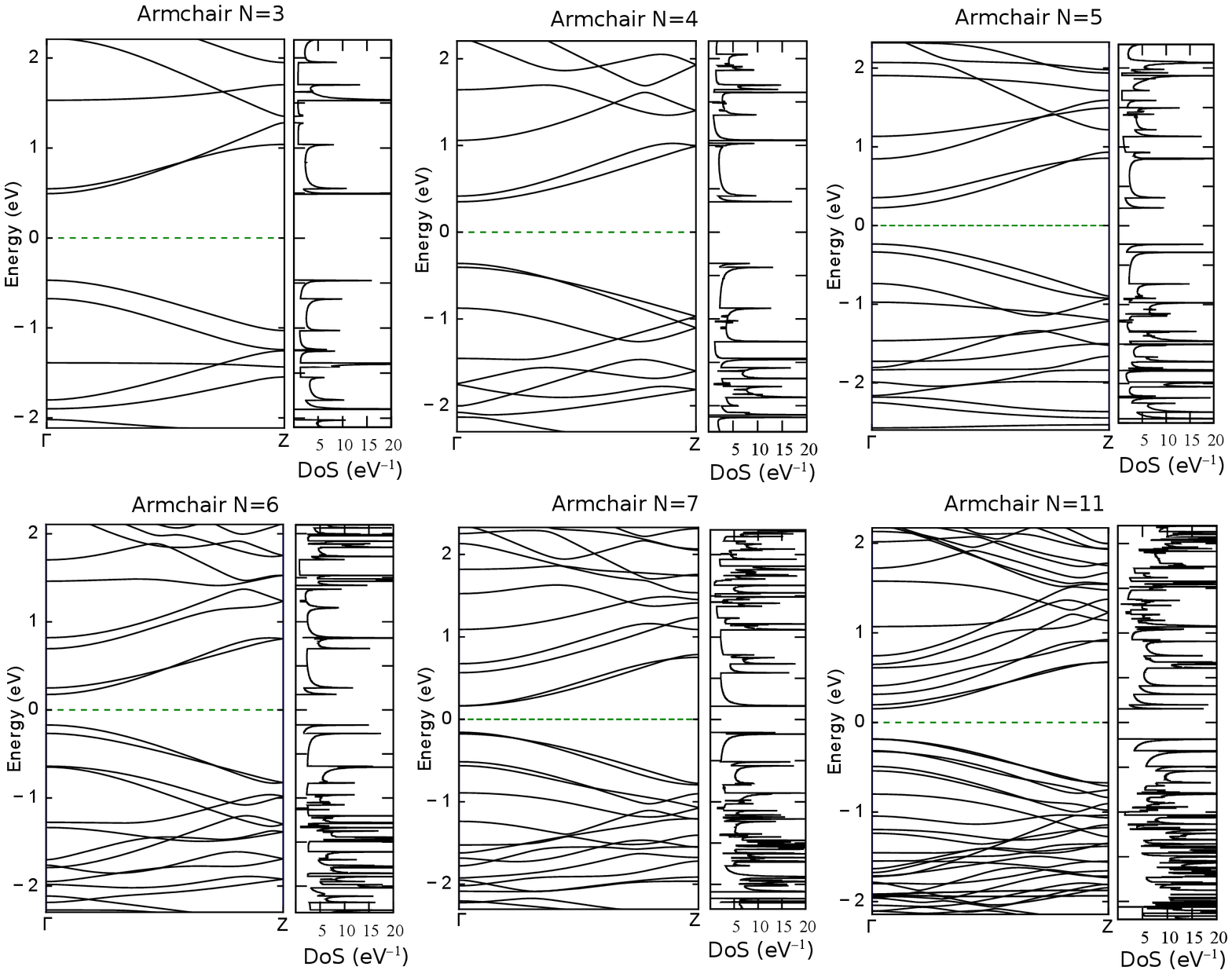}\\
	\caption{Bandstructure and density of states for AGRNRs.}
	\label{fig:bg_armchair}
	
	\
	
	\includegraphics[width=0.8\linewidth]{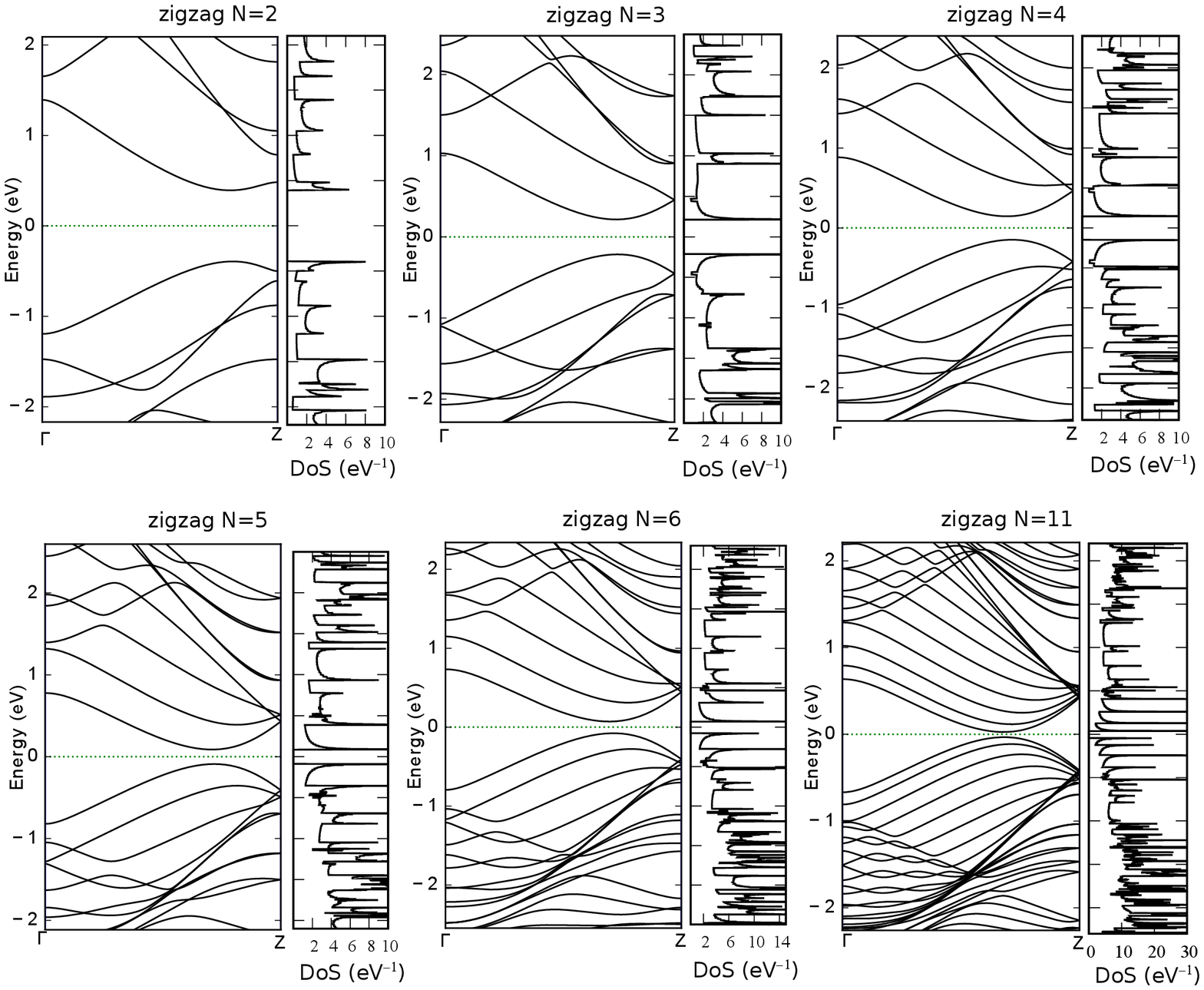}\\
	\caption{Bandstructure and density of states for ZGRNRs.}
	\label{fig:bg_zigzag}
\end{figure}

We consider AGRNRs with the number of dimeric lines $N=3, 4, 5, 6, 7, 11$ and ZGRNRs with the number of zigzag chains $N=2,3,4,5,6,7,11$. In Figure~\ref{fig:bg_armchair}, bandstructure and density of states (DOS) for AGRNRs are demonstrated. The dashed line indicates the Fermi level. One can see that all considered AGRNRs are direct-gap semiconductors with a narrow bandgap. The minimum of the conduction band and the maximum of the valence band are located at the $\Gamma$ point. The largest gap is observed for the narrowest AGRNR (table~\ref{tab:stab}). With an increase of the nanoribbon width, the bandgap decreases.

The bandstructures of ZGRNRs are shown in Figure~\ref{fig:bg_zigzag}.  Similar to armchair ribbons, they are direct-gap semiconductors with a narrow bandgap. However, the minimum gap is located between $\Gamma$ and Z points, closer to the latter. With an increase of the nanoribbon width, the bandgap decreases (see Table~\ref{tab:stab}). Comparing armchair and zigzag ribbons with the same value of $N$, one can see that AGRNR has a larger bandgap than ZGNRN. As the ribbon width increases, this difference disappears, and $E_{g}$ tends to 0.051~eV corresponding to the graphenylene sheet.

\section{Optical properties}

To study the optical properties of GRNR, we orient ribbons along $z$-axis in such a way that the ribbon plane is perpendicular to the $x$-axis.
The optical conductivity, the complex refractive index, and the absorption coefficient are calculated by the following formulas
\begin{equation}
\sigma=-i\omega\epsilon_{0}\chi(\omega),\quad
\textit{n}+\textit{i}\kappa=\sqrt{1+i\frac{\sigma}{d\epsilon_{0}\omega}},
\quad			
\alpha=2\frac{\omega}{c}\kappa.
\end{equation}

The dielectric susceptibility $\chi(\omega)$ is determined from the Kubo-Greenwood formalism implemented in the Quantum ATK package. To determine the effective thickness $d$ of the graphenylene layer,  the two-layer structure is optimized and the optimal distance between the sheets of 3.55~\AA is used.

\begin{figure}[!h]
	\centering
	\includegraphics[width=0.8\linewidth]{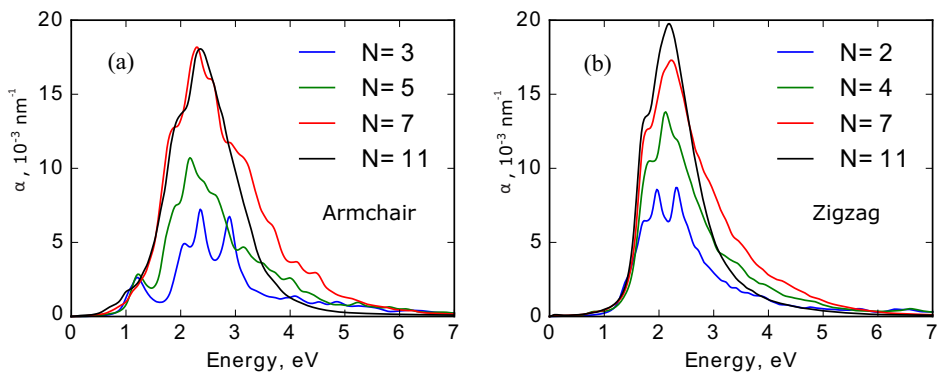}
	\caption{Absorption coefficient of AGRNRs (a) and ZGRNRs (b).}
	\label{ko4_abs} 
\end{figure}
\begin{figure}[!h]
	\centering
	\includegraphics[width=0.8\linewidth]{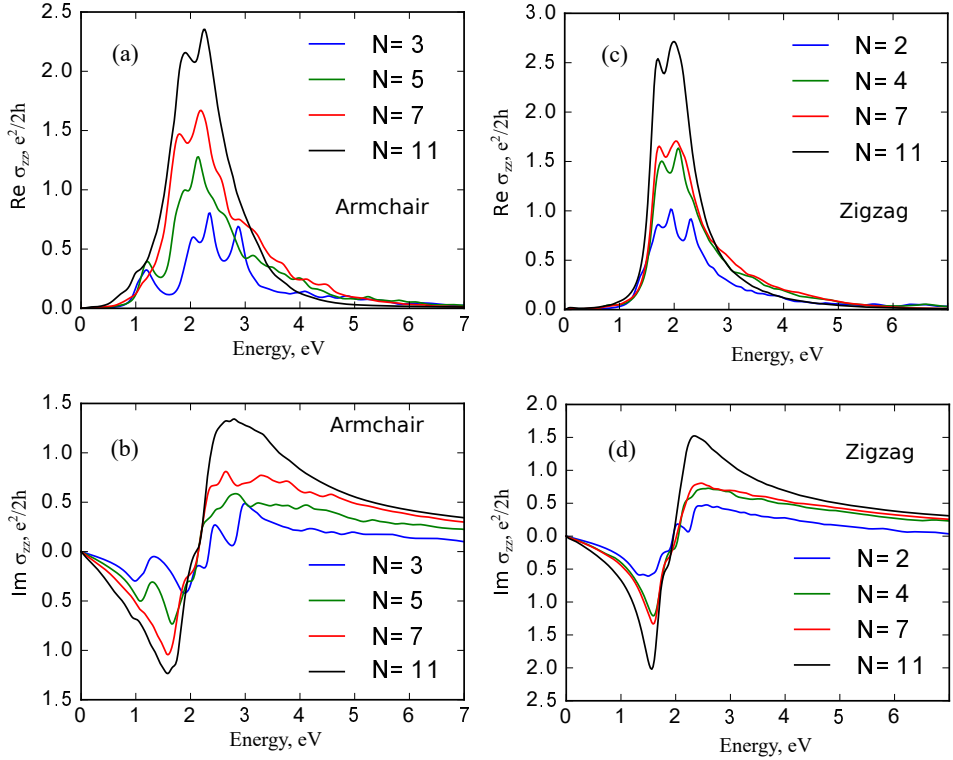}
	\caption{Optical conductivity of AGRNRs (a,b) and ZGRNRs (c,d).}
	\label{sigma_zig} 
\end{figure}
\begin{figure}[!h]
	\centering
	\includegraphics[width=0.8\linewidth]{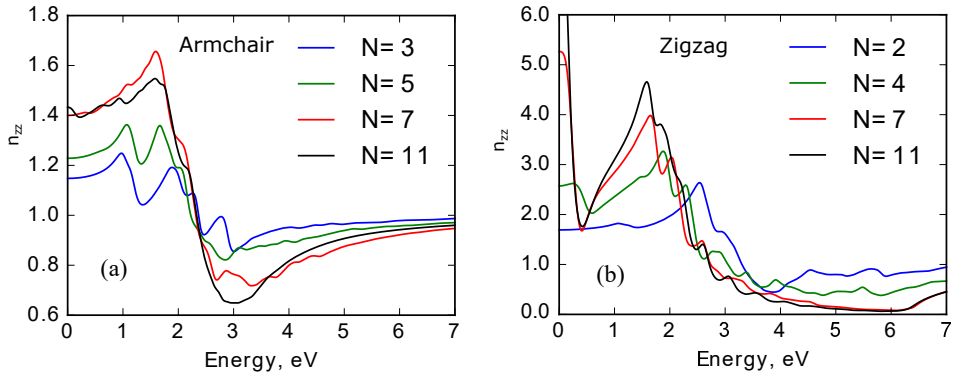}
	\caption{Refractive index of  AGRNRs (a) and ZGRNRs (b)}
	\label{ko4_Ren_zig} 
\end{figure}

The absorption spectrum, complex optical conductivity and refractive index of the ZGRNR (N = 2, 4, 6, 11) and AGRNR (N = 3, 5, 7, 11) are shown in Figures~\ref{ko4_abs} -- \ref{ko4_Ren_zig}. The imaginary part of the refractive index is represented by the absorption coefficient and the real part of the conductivity. In the present paper, waves with polarization along nanoribbons are considered. Obviously, the $x$-components of the quantities are equal to zero. The transverse polarization along the $y$ axis is not considered. Due to the depolarization effect, the absorption of waves is strongly suppressed, therefore, the nanoribbon is almost transparent for waves with such polarization. 


The imaginary part of the optical conductivity for all ribbons takes negative values at low photon energies. This indicates the presence of plasma reflection in this range~(Figure~\ref{sigma_zig}).

For all GRNRs, an anomalous dispersion region is observed, and it covers the entire visible range (Figure \ref{ko4_Ren_zig}). Their centers coincide with the main absorption bands. With increasing ribbon width, this region shifts towards lower energies. Starting with $N = 4$ for ZGRNR and with $N = 5$ for AGRNR, the region of anomalous dispersion also appears in the low-frequency region.


\section{Thermoelectric properties}

The thermoelectric efficiency of the material can be characterized by the figure of merit:
\begin{displaymath}
ZT=\frac{S^{2}GT}{\lambda},
\end{displaymath}
where $S$ is the Seebeck coefficient, $G$ is the electric conductivity, $T$ is the absolute temperature, and $\lambda$ is the thermal conductivity, which is equal to the sum of electron $\lambda_{e} $ and phonon $\lambda_{ph} $ thermal conductivity.
To calculate the transfer coefficients, the method of nonequilibrium Green functions (NEGF) and the method of nonequilibrium molecular dynamics are combined.
Electronic transport properties are calculated using the NEGF-formalism implemented in the Quantum ATK package. Here, the common approach is employed, in which the central part of a ribbon is connected to the semi-infinite left and right parts.

\begin{figure}[!h]
	\centering
	\includegraphics[width=0.9\linewidth]{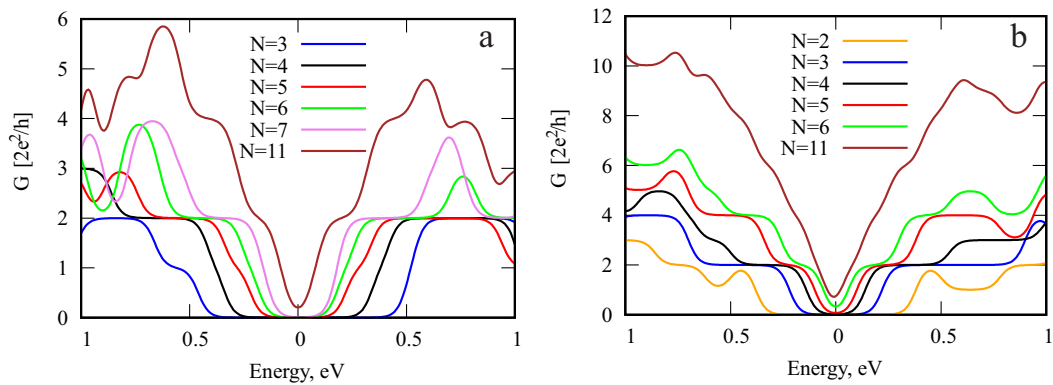}
	\caption{Electrical conductivity: a -- AGRNR, b -- ZGRNR.}
	\label{conductance}
\end{figure}

Figure \ref{conductance} shows the calculation results for the conductivity and Seebeck coefficient of the armchair and zigzag GRNR depending on the chemical potential. Calculations are performed for the room temperature $T=300$~K. Due to the broadening of energy levels ($T \ne0 $), the conductivity graph does not have clear steps. There is also no symmetry between the conductivity of p- and n-type graphenylene semiconductors. For example, for 3-AGRNR the first conduction channel is 1 for the p-type impurity, and for the n-type it is 2. This is due to the different number of bands at energies corresponding to the opening of the channel. In the vicinity the Fermi level, the conductivity is 0, the ribbons are semiconducting. With increasing ribbon width, the energy range with zero conductivity decreases that corresponds to a decrease of the bandgap. For the widest considered ribbons (11-AGRNR, 6-ZGRNR, and 11-ZGRNR), the conductivity at zero energy is nonzero despite the fact that they are semiconductors. This is also related to the broadening of energy levels and, as a consequence, the overlap of the $G(E)$ function for $p$ and $n$-type  channels with a smooth change in conductivity.

\begin{figure}[!h]
	\centering
	\includegraphics[width=0.9\linewidth]{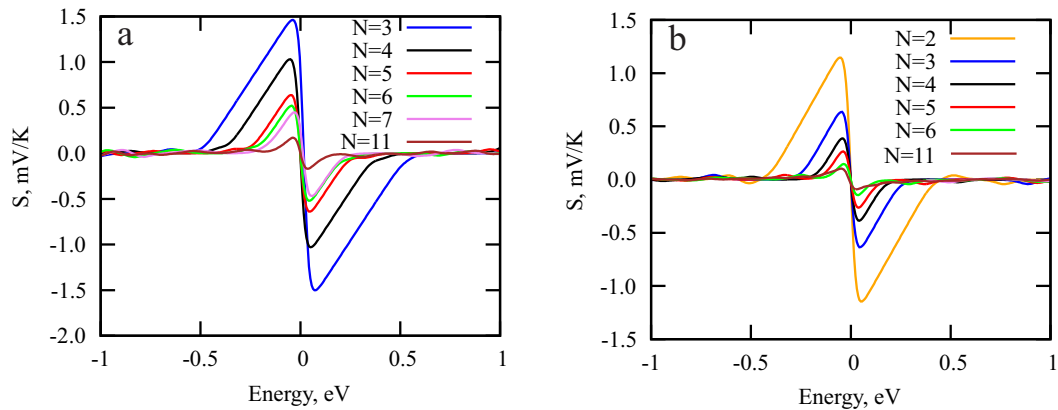}
	\caption{Seebeck coefficient: a -- AGRNR, b -- ZGRNR.}
	\label{Seebeck coeff}
\end{figure}

The Seebeck coefficient for ribbons of both types has the largest absolute value near the Fermi level and increases for narrow ribbons (Fig. \ref{Seebeck coeff}). It takes a maximum value of 1.46~mV/K for 3-AGRNR. 

\begin{figure}[!h]
	\centering
	\includegraphics[width=0.9\linewidth]{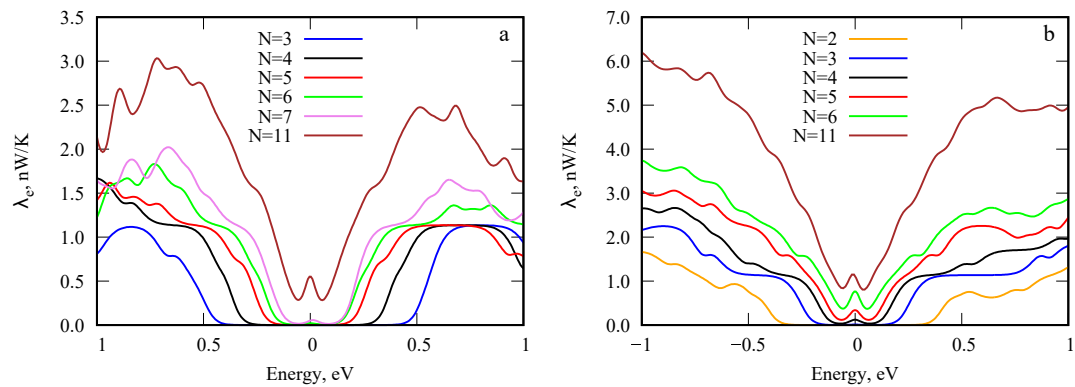}
	\caption{Electronic thermal conductivity: a -- AGRNR, b -- ZGRNR.}
	\label{termal electron conductance}
\end{figure}
\begin{figure}[!h]
	\centering
	\includegraphics[width=0.65\linewidth]{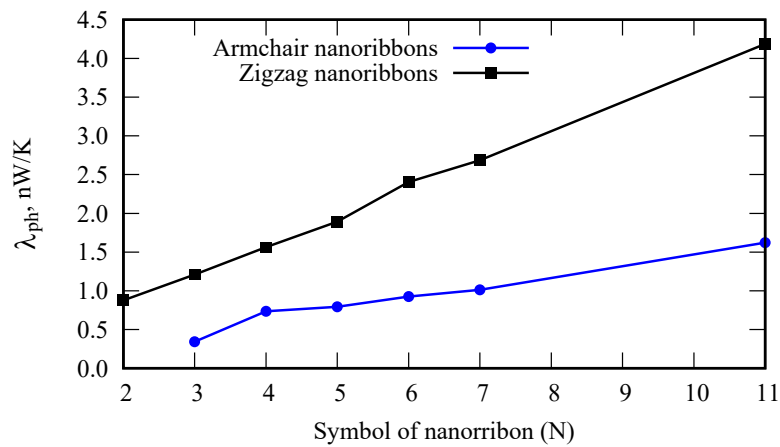}
	\caption{Phonon thermal conductivity.}
	\label{termal phonon conductance}
\end{figure}

To study the heat-conducting properties, the electron (Figure \ref{termal electron conductance}) and phonon (Fig. \ref{termal phonon conductance}) thermal conductivity coefficients are calculated for the studied nanoribbons. Electronic thermal conductivity is asymmetric with respect to the Fermi level. Phonon thermal conductivity increases for ribbons with larger width, because the number of phonon modes increases. Moreover, the thermal conductivity of ZGRNR is greater than the thermal conductivity of AGRNR. This is due to different scattering frequencies at the zigzag and armchair edges. The phonon contribution to the total thermal conductivity is comparable with the electronic one for both ZGRNR and AGRNR.

\begin{figure}[!h]
	\centering
	\includegraphics[width=0.9\linewidth]{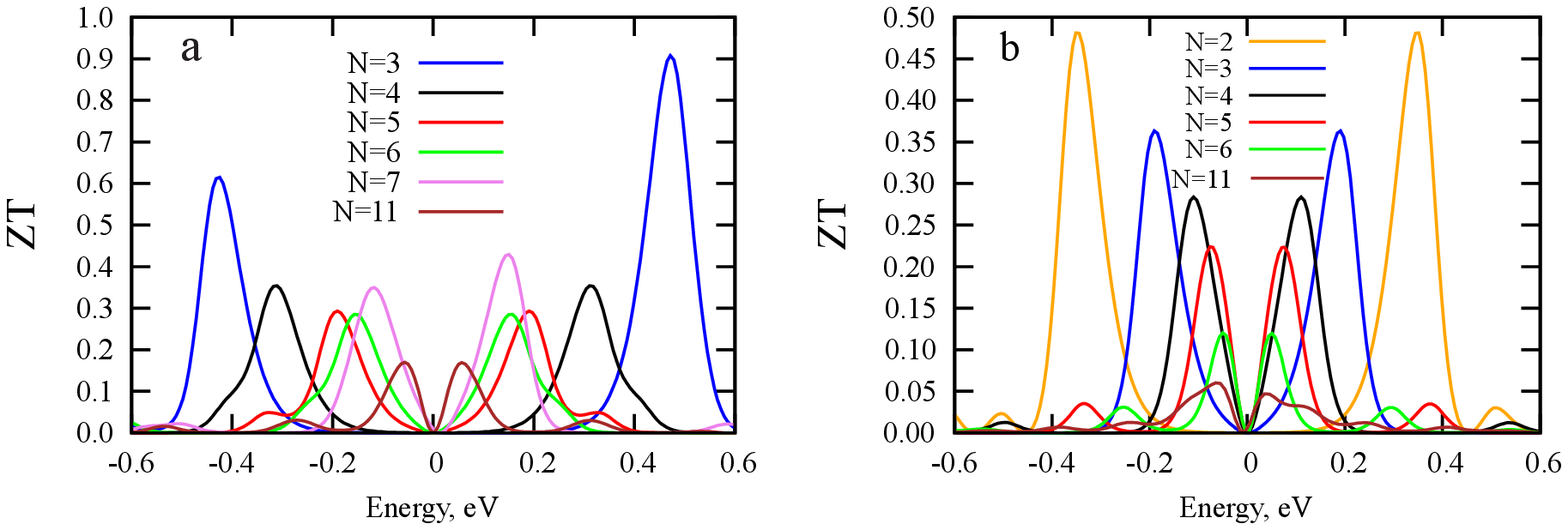}
	\caption{Thermoelectric figure of merit: a -- AGRNR, b -- ZGRNR.}
	\label{ZT}
\end{figure}

In conclusion, we calculated the dependencies of the dimensionless thermoelectric figure of merit at room temperature on energy for both types of ribbons (Figure \ref{ZT}). One can see that for ZGRNR the $ZT$ vs $E$ graph is symmetric with respect to $E=0$. For AGRNR there is no such symmetry, and it can be seen that n-type doping is more profitable to obtain a material with higher thermoelectric figure of merit. Narrow ribbons have the highest values of $ZT$. For 3-AGRNR $ZT=0.91 $, and for 2-ZGRNR $ZT=0.48 $. This is not enough to consider graphenylene ribbons as promising thermoelectric materials.

\section{Conclusion}

Using DFT, NEGF and molecular dynamics method, the electrical, optical, and thermoelectric properties of graphenylene nanoribbons have been calculated. The considered zigzag and armchair ribbons are direct-gap semiconductors with a narrow bandgap. For the increased ribbon width, the energy bandgap decreases and tends to the value typical for a graphenylene sheet that is a two-valley direct-gap semiconductor. It is known that the presence of two valleys in the bandstructure leads to $N$-shaped $IV$-plots comprising a section with negative differential resistance. For this reason, if a graphenylene ribbon of sufficient width is cut, a semiconductor material with similar properties can be obtained. We have shown that the graphenylene bandgap can be effectively controlled within a sufficiently wide range by the applied tensile stresses. Such materials can be used in various elements of nanoelectronics. 

The calculated absorption coefficient shows that graphenylene ribbons effectively absorb the energy of electromagnetic waves in the visible range, which leads to a significant increase in photoconductivity. Thus, the graphenylene nanoribbons are expected to find application in photodetectors operating in the visible range.

Using NEGF, DFT and molecular dynamics method, we have calculated the transport coefficients and thermoelectric figure of merit for GRNRs.
The maximum value of the thermoelectric figure of merit obtained for graphenylene ribbons is 0.91 for 3-AGRNR. 

\section*{Acknowledgment}

Renat Sibatov acknowledges the financial support from the Russian Science Foundation (project 19-71-10063).

\end{document}